\documentclass[11pt,a4paper]{article}
\pdfoutput=1
\usepackage{graphicx}
\usepackage{amssymb}
\usepackage{amsfonts}
\usepackage{amsmath}
\usepackage{mathrsfs}

\usepackage{hyperref}

\newskip\humongous \humongous=0pt plus 1000pt minus 1000pt

\newif\ifdtup

\catcode`\@=11

\@addtoreset{equation}{section}
\def\theequation{\thesection.\arabic{equation}}

\def\@normalsize{\@setsize\normalsize{15pt}\xiipt\@xiipt
\abovedisplayskip 14pt plus3pt minus3pt%
\belowdisplayskip \abovedisplayskip
\abovedisplayshortskip \z@ plus3pt%
\belowdisplayshortskip 7pt plus3.5pt minus0pt}

\def\small{\@setsize\small{13.6pt}\xipt\@xipt
\abovedisplayskip 13pt plus3pt minus3pt%
\belowdisplayskip \abovedisplayskip
\abovedisplayshortskip \z@ plus3pt%
\belowdisplayshortskip 7pt plus3.5pt minus0pt
\def\@listi{\parsep 4.5pt plus 2pt minus 1pt
     \itemsep \parsep
     \topsep 9pt plus 3pt minus 3pt}}

\relax

\catcode`@=12

\catcode`\@=11

\def\section{\@startsection{section}{1}{\z@}{3.5ex plus 1ex minus
   .2ex}{2.3ex plus .2ex}{\large\bf}}

\def\thesection{\arabic{section}}
\def\thesubsection{\arabic{section}.\arabic{subsection}}

\def\appendix{\setcounter{section}{0}
 \def\thesection{Appendix \Alph{section}}
 \def\thesubsection{\Alph{section}.\arabic{subsection}}
 \def\theequation{\Alph{section}.\arabic{equation}}}

\def\SymBoxes#1#2#3#4{\newdimen\un@t \un@t#3%
\raisebox{#1}{\rule{#2\un@t}{#4}\hskip-#2\un@t
\@tempdimb\un@t \advance\@tempdimb by-#4\@tempcntb#2\relax%
\@whilenum{\@tempcntb>0}\do{
\rule{#4}{\un@t}\hskip\@tempdimb \advance\@tempcntb by\m@ne}%
\hskip-#2\un@t \rule[\un@t]{#2\un@t}{#4}%
\rule[\un@t]{#4}{#4}\hskip-#4
\rule{#4}{\un@t}}\hskip-#4}                

\begin{document}

\newcommand{\beq}{\begin{equation}}
\newcommand{\eeq}{\end{equation}}
\newcommand{\bea}{\begin{eqnarray}}
\newcommand{\eea}{\end{eqnarray}}
\newcommand{\beas}{\begin{eqnarray*}}
\newcommand{\eeas}{\end{eqnarray*}}
\newcommand{\defi}{\stackrel{\rm def}{=}}
\newcommand{\non}{\nonumber}
\newcommand{\bquo}{\begin{quote}}
\newcommand{\enqu}{\end{quote}}
\renewcommand{\(}{\begin{equation}}
\renewcommand{\)}{\end{equation}}
\def\IZ{{\mathbb Z}}
\def\IR{{\mathbb R}}
\def\IC{{\mathbb C}}
\def\IQ{{\mathbb Q}}

\def\g{\gamma}
\def\m{\mu}
\def\n{\nu}
\def\b{\beta}

\def\a{{\textsl a}}

\def\CM{{\mathcal{M}}}
\def\dCM{{\left \vert\mathcal{M}\right\vert}}

\def \d{\textrm{d}}
\def \p{\partial}

\def \Pf{\rm Pf\ }

\def \pr{\prime}

\def\Tr{ \hbox{\rm Tr}}
\def\half{\frac{1}{2}}

\def \eqn#1#2{\begin{equation}#2\label{#1}\end{equation}}
\def\de{\partial}
\def\Tr{ \hbox{\rm Tr}}
\def\H{ \hbox{\rm H}}
\def\HE{ \hbox{$\rm H^{even}$}}
\def\HO{ \hbox{$\rm H^{odd}$}}
\def\K{ \hbox{\rm K}}
\def\Im{ \hbox{\rm Im}}
\def\Ker{ \hbox{\rm Ker}}
\def\const{\hbox {\rm const.}}
\def\o{\over}
\def\im{\hbox{\rm Im}}
\def\re{\hbox{\rm Re}}
\def\bra{\langle}\def\ket{\rangle}
\def\Arg{\hbox {\rm Arg}}
\def\Re{\hbox {\rm Re}}
\def\Im{\hbox {\rm Im}}
\def\exo{\hbox {\rm exp}}
\def\diag{\hbox{\rm diag}}
\def\longvert{{\rule[-2mm]{0.1mm}{7mm}}\,}
\def\a{{\textsl a}}
\def\dag{{}^{\dagger}}
\def\tq{{\widetilde q}}
\def\p{{}^{\prime}}
\def\W{W}
\def\N{{\cal N}}
\def\hsp{,\hspace{.7cm}}
\newcommand{\C}{\ensuremath{\mathbb C}}
\newcommand{\Sp}{\ensuremath{\mathbb S}}
\newcommand{\Z}{\ensuremath{\mathbb Z}}
\newcommand{\R}{\ensuremath{\mathbb R}}
\newcommand{\rp}{\ensuremath{\mathbb {RP}}}
\newcommand{\cp}{\ensuremath{\mathbb {CP}}}
\newcommand{\vac}{\ensuremath{|0\rangle}}
\newcommand{\vact}{\ensuremath{|00\rangle}}
\newcommand{\oc}{\ensuremath{\overline{c}}}
\newcommand{\sgn}{\mathop{\mathrm{sgn}}}

\def\M{\mathcal{M}}
\def\F{\mathcal{F}}
\def\d{\textrm{d}}

\def\eps{\epsilon}

\begin{flushright}
\end{flushright}

\vspace{-.5truecm}
\begin{center}
{\Large \textbf{Can Froissart Bound Explain\\[7pt] Hadron Cross-Sections at High Energies?}}
\end{center}
\vspace{6pt}
\begin{center}
{\large\textsl{Anatoly Dymarsky $^{a}$}\\}
\vspace{25pt}
\textit{ $^a$ Skolkovo Institute of Science and Technology,\\
Novaya St.~100, Skolkovo, Moscow Region, Russia, 143025} \vspace{6pt}
\end{center}

\vspace{6pt}
\begin{center}
\textbf{Abstract}
\end{center}
\vspace{-.1cm}
Experimentally observed slow growth of hadron cross-sections at high energies is a very intriguing but poorly understood property of QCD. It is tempting to explain the slow growth by saturation of Froissart bound or another similar universal mechanism.  We reconsider derivation of Froissart bound in QCD in chiral limit and argue it can not justify experimentally observed behavior. Although the conventional Froissart-Martin bound should impose non-trivial constraint on the growth of hadron cross-sections, because of the small value of pion masses it will become restrictive only at currently unaccessible center-of-mass energies exceeding $10^5-10^6$ GeV.

\vspace{6pt}

\renewcommand{\thefootnote}{\arabic{footnote}}

\section{Introduction}
\begin{figure}[t]
\includegraphics[width=1\textwidth]{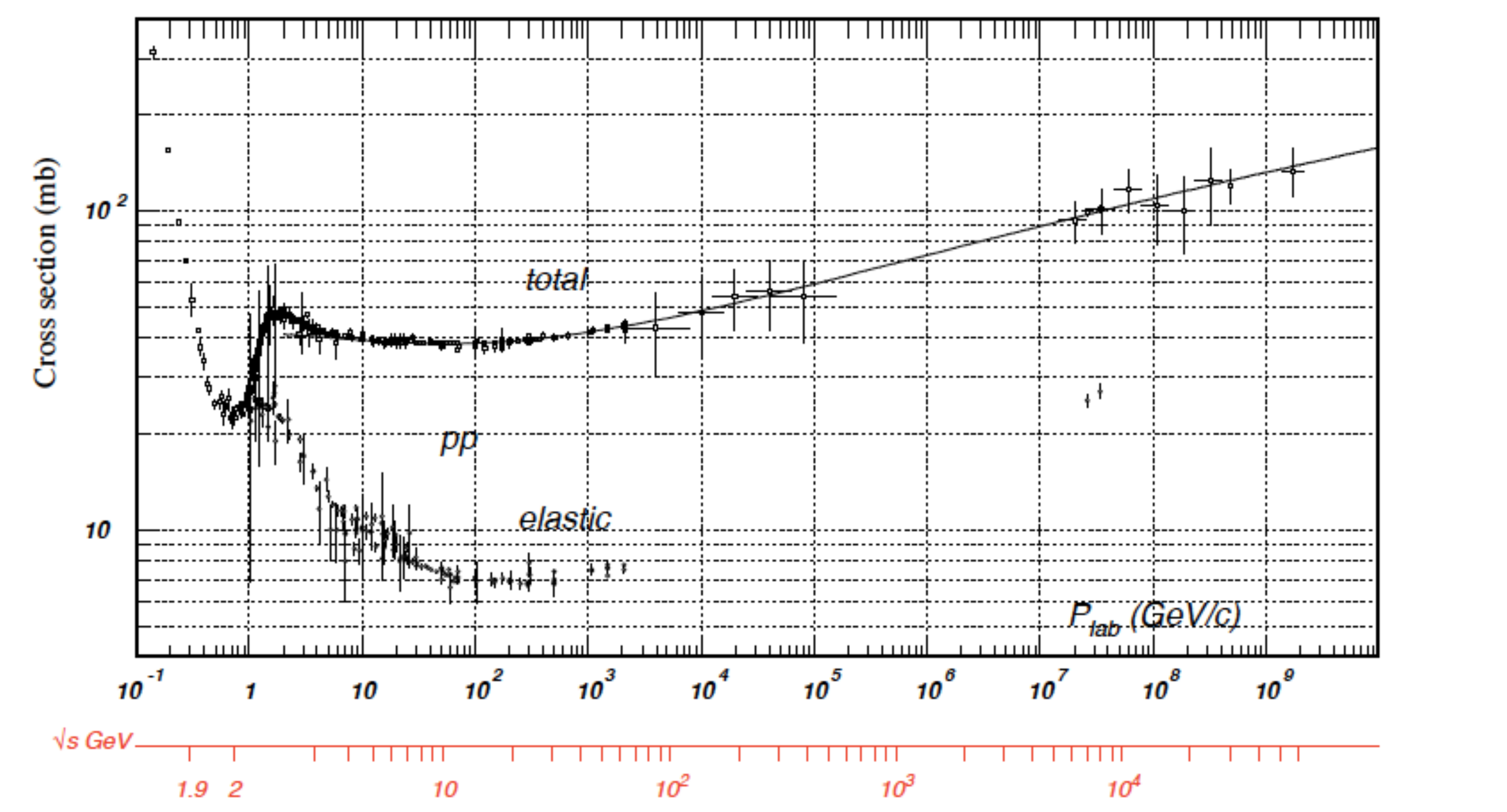}
\label{fig:pp}
\caption{Proton-proton total cross-section as a function of center-of-mass energy $\sqrt{s}$. The plot is taken from \cite{PDG}.}
\end{figure}

Experimentally observed slow growth of hadron cross-sections at high energies is an intriguing phenomenon calling for an explanation within the framework of quantum chromodynamics. For concreteness in this paper we  focus on total cross-section of proton-proton scattering. It changes only about $2.5$ times, roughly from  40 mb to 120 mb, while the center-of-mass energy  changes by several orders of magnitude from 10 to $10^5$ GeV \cite{PDG}, see Fig.~\ref{fig:pp}. This very slow  growth can be approximated with a good accuracy for the center-of-mass energies exceeding 100 GeV with the following fit (see  \cite{PDG}),
\bea
\label{log2}
\sigma_{pp}\simeq34.71+0.2647 \log^2\left({s\over 16}\right)\ .
\eea
Here $\sigma_{pp}$ is a total proton-proton cross-section measured in mb, while  the Mandelstam variable $s$ is measured in GeV$^2$.
Other hadron cross-sections, e.g.~of proton-anti-proton scattering, exhibit similar slow  growth behavior \cite{PDG}. 

Because the theory in question is strongly coupled,  high-energy behavior of $\sigma_{pp}$ can not be immediately deduced from QCD dynamics. At the same time ``logarithm square'' growth of \eqref{log2} looks temptingly similar to  the  behavior of the upper bound on the total cross-section proposed by Froissart in \cite{Froissart},
\bea
\label{Froissart}
\sigma_F=\sigma_0  \log^2\left({s\over s_0}\right)\ .
\eea 
The constant $\sigma_0$ is inverse-proportional to the nearest location of the branch-cut singularity in $t$-channel, $\sigma_0\sim 1/t_0$, and $s_0$ is some appropriate constant. Any total cross-section $\sigma$ in a given gapped theory may not exceed $\sigma_F$ when energy $s$ is sufficiently large. It looks deceptively simple to declare the particular form of \eqref{log2} to be a consequence of saturation of Froissart bound \eqref{Froissart}.

In the context of QCD the constant $\sigma_0$ was fixed in \cite{Martin:1964zz} by Martin and $\L$ukaszuk leading to Froissart-Martin bound $\sigma^{FM}_0={\pi/m^2_{\pi}}\simeq$ 63.32 mb. (The value of $s_0$ was recently fixed to be $s_0=m^2_{\pi}\sqrt{2}/(17 \pi^{3/2})$ \cite{Martin:2013xcr}, see also \cite{Martin:2009pt, Wu:2010sd} for related progress.) The two orders of magnitude difference between $\sigma^{FM}_0$ and the coefficient in front of $\log^2(s)$ in \eqref{log2} is a clear indication that Froissart-Martin bound is currently not saturated and thus can not justify experimentally observed behavior of $\sigma_{pp}$. 

Several authors suggested that the huge discrepancy between the actually observed slope of $\log^2(s)$ in \eqref{log2} and the prediction of the Froissart-Martin bound $\sigma^{FM}_0$ is due to pseudo-Goldstone nature of pions. In chiral limit $m_\pi\rightarrow 0$ naive Froissart bound becomes obsolete, but it was argued there is another improved bound of the form \eqref{Froissart} with finite value of $\sigma_0$. Since large impact parameters correspond to a small invariant mass of the pion pair, and low energy Goldstone bosons decouple, it was argued in \cite{KS} that $\sigma_0$ would be determined by the strong interaction scale of $M\sim$1 GeV, $\sigma_0\sim 1/M^2$. A similar conclusion was also reached in \cite{Giordano:2013iga,Giordano:2014cia} and in \cite{Greynat:2013cta,Greynat:2013zsa}, where $\sigma_0$ was calculated to be given by some combination of $\rho$-meson mass and $f_{\pi}$. 

In section 2 we reconsider derivation of Froissart bound in presence of massless or very light (pseudo)-Goldstone bosons and conclude that the asymptotic form \eqref{Froissart} with $\sigma_0\sim 1/M^2$, $M\sim$ 1 GeV is unwarranted. Furthermore, even if the resulting value of $\sigma_0$ would approximately match the coefficient in front of $\log^2(s)$ from \eqref{log2}, still the behavior of \eqref{log2} for $10^2<\sqrt{s}<10^5$ GeV could not be justified by saturation of \eqref{Froissart}. This is because the bound \eqref{Froissart}, if holds,  has to apply whenever $\sigma_{pp}(s)\gg \sigma_0$ (this point is justified in section 2 below), but presence of the constant term in \eqref{log2} precludes saturation of \eqref{Froissart} for $10^2<\sqrt{s}<10^5$ GeV. We further elaborate on this point in Summary.

To summarize, our first conclusion is that the slow growth of the  proton-proton total cross-section experimentally observed for energies smaller than $10^5$ GeV can not be attributed to saturation of the appropriately modified Froissart bound. Still, it would be interesting to understand if a similar logic can help predict the asymptotic behavior of proton-proton total cross-section when $s\rightarrow \infty$. Let us note that we can not take for granted that the form \eqref{log2} will persist much longer beyond currently accessible scales. Indeed there is some evidence  $\sigma_{pp}$ may grow faster than the ``logarithm square'' growth of \eqref{log2} \cite{Azimov, Menon}.
This conclusion is in agreement with an alternative fit of experimental data \cite{Landshoff:2008cz},
\bea
\label{regge}
\sigma_{pp}\simeq 21.7\, s^{0.0808}\ ,
\eea 
which agrees well with \eqref{log2} for $10^2<\sqrt{s}<10^6$ GeV but grows much faster when $s\rightarrow \infty$. Clearly \eqref{regge} would start violating \eqref{Froissart} with any $\sigma_0,s_0$ when $s$ is significantly large.\footnote{Here we implicitly assume that scattering amplitude as a function of energy grows not faster than a polynomial. This assumption may be unjustified allowing $\sigma\sim s^\alpha$ at all energies \cite{Azimov,Azimov2}.} As was mentioned above, if  Froissart-Martin bound holds, it must apply when $\sigma_{pp}$ becomes significantly larger than  $\sim 2\sigma_0^{FM}\simeq 130$ mb which is expected to happen at energies of order $\sqrt{s}\sim 10^5-10^6$ GeV.  Obviously we can not say if $\sigma_{pp}$ will eventually saturate the bound but we expect that $\sigma_{pp}$ will exhibit its true asymptotic behavior starting at or below this scale. 

\section{Froissart Bound Revisited}
Let us briefly remind the reader main steps leading to the derivation of Froissart bound. We consider scattering of two identical particles at center-of-mass energies $\sqrt{s}$ much larger than their mass. 
Using optical theorem the total cross-section can be expressed through the imaginary part of forward scattering amplitude 
\bea
\label{sigmatot}
\sigma_{\rm total}={4\pi \over s}{\Im\, A(s,\cos\theta=1)}\ .
\eea
The $2\rightarrow 2$ elastic scattering amplitude $A(s,\cos\theta)$ can be decomposed into partial waves with help of the Legendre polynomials  $P_l$,
\bea
\label{partialwaves}
A(s, \cos\theta)=\sum_{l=0}^\infty a_l(s) (2l+1) P_l(\cos\theta)\ .
\eea
The upper bound on \eqref{sigmatot} is a result of the inequality $\Im\, a_l\le | a_l(s)|<1$, which is a consequence of unitarity, and a bound on $\Im\, a_l\le| a_l(s)|$  for large $l$ following from analyticity of $A(s,x)$ with respect to $x$. Rewriting $x$ through Mandelstam variables $x=\left(1+{2t\over s}\right)$ we conclude $A(s,x)$ as a function of $x$ must have branch-cuts along real axis for $|x|>z_0$ where $z_0=1+2t_0/s$ and $t_0$ is the energy squared of the lightest state appearing in $t$-channel. Assuming these branch-cuts are the only singularities of $A(s,x)$,\footnote{This is a technical assumption which we believe can be avoided. Thus, \cite{Azimov2} derives Froissart bound using only analyticity of $A$ inside Lehman ellipse, without making any explicit assumptions about structure of the singularities of $A(s,x)$.}  
one can express $A(s,x)$ using dispersion relation 
\bea
\label{disp}
A(s,x)&=&\int^\infty_{z_0} {dz\over z-x}\, \rho(s,z)\ , \\ \nonumber
2 i\pi\rho(s,z)&=&A(s,z+i\epsilon)-A(s,z-i\epsilon)+A(s,-z-i\epsilon)-A(s,-z+i\epsilon)\ .
\eea
The dispersion relation  \eqref{disp} may be divergent. In such a case corresponding integral  should be regularized by a number of subtractions. 
Consequently $a_l(s)$ can be expressed through the branch-cut jump function $\rho$ and Legendre function of the second kind $ Q_l(z)$ as follows (for $l$ large enough this integral is convergent and does not require subtractions)
\bea
a_l(s)=\int_{z_0}^\infty Q_l(z) \rho(s, z)\ .
\eea
Using asymptotic from of $Q_l(z)$  for large $l$  derived in appendix this integral can be estimated from above with help of Laplace method. First, let us assume $|\rho(s,z)|$ near $z\rightarrow 1^+$ can be approximated, or bounded from above, by some function ${\mathcal A}(s)$, $|\rho(s,z)|\le {\mathcal A}(s)$. Then 
\bea
\label{estimate}
\a_l(s)
&\equiv&{\mathcal A}(s)\int_{z_0}^\infty Q_l(z) \lesssim {\mathcal A}(s)
\left\{
\begin{array}{l}
{1\over l^2}\quad \ \, \ \   \qquad\qquad \ ,\quad  z_0-1\ll {1\over 8l^2} \\
{\sqrt{\pi} \sqrt{1-e^{-2\alpha_0}}\over  2 l^{3/2}}e^{-\alpha_0 l}\ ,\quad z_0-1\gtrsim {1\over 8l^2}
\end{array}\right. \,\\
|a_l(s)|&\le & \a_l(s) \ . \label{const}
\eea
Integral \eqref{estimate} is a monotonically deceasing function of $l$. Starting from some $l=L$  constraint \eqref{const} will become better than the unitarity constraint $|a_l(s)|<1$. Following \cite{Azimov2} we introduce the combination $Y=\alpha_0 L$, where 
\bea
\label{alpha}
\alpha_0=\log(z_0+\sqrt{z_0^2-1})\ ,
\eea
and notice that in the limit $t_0/s\rightarrow 0$ conditions $Y\gg 1$ and $z_0-1\gg 1/L^2$ are the same. 

Assuming that ${\mathcal A}(s)$ grows with $s$, at some point ${\mathcal A}(s)$ will become large enough such that corresponding $Y\sim \log{\mathcal A}(s) \gg 1$. To find a bound on \eqref{sigmatot} we split the sum \eqref{partialwaves} into two parts, from $0$ to $L-1$ and from $L$ to infinity and use unitarity bound $|a_l|\le1$ and \eqref{const} correspondingly. Keeping only leading terms we arrive at (see e.g.~\cite{Azimov2} for a similar calculation) 
\bea
\label{fb}
\sigma_{\rm total}\le {4\pi \over s}(L^2+2L/\alpha_0-1/\alpha_0^2)= \sigma_0 ((Y+1)^2-2)\ ,\quad \sigma_0\equiv{\pi\over t_0}\ .
\eea
Notice that by assumption $Y\gg 1$ and hence \eqref{fb} is always positive.  
Common lore predicts ${\mathcal A}(s)\sim s^N$ for large $s$, leading to $\sigma_0=\pi N^2/t_0$ and asymptotic form \eqref{log2}. 
Taking ${\mathcal A}(s)\sim s^2$ and $t_0=4m_\pi^2$ we recover the Froissart-Martin bound for hadron scattering\footnote{Strictly speaking the lightest singularity in $t$-channel is a one pion pole at $t_0=m_\pi^2$. It can be shown though that the contribution of pole singularities goes to zero when $s\rightarrow \infty$, see \cite{Azimov2}. Hence the leading singularity contributing to the Froissart bound is a two-pion state $t_0=4m^2_\pi$.}
\bea
\label{FM}
\sigma_{\rm total}\lesssim {\pi \over m_\pi^2}\log^2(s/s_0)\ .
\eea

Different assumptions about $\rho$ may lead to different $1/l$ suppression of $\a_l$ for large $l\gg \alpha_0^{-1}$ which will result in different $Y$-independent term in \eqref{fb}. In any case this constant term is always of order $1$ and can not justify the asymptotic form $\sigma_F=\sigma_0 \log^2(s/s_0)+\sigma_1$ with $\sigma_0$ and $\sigma_1$ being different by several orders of magnitude as in \eqref{log2}.

An interesting but rarely discussed possibility is when $\sqrt{s}$  is much larger than the masses of external particles and $\sqrt{t_0}$, but ${\mathcal A}(s)$ is not too large such that corresponding $Y$ is of order or smaller than one. Then for $l\le L$, $\a_l={\mathcal A}(s)/l^2$ and $L\simeq {\mathcal A}^{1/2}(s)$. Once again we split the sum \eqref{const} into two parts, but now the sum from $l=L$ to infinity is tricky. It starts as $\sim\sum (2l+1)/l^2$ for $L<l\lesssim\alpha_0^{-1}$ and continues as $\sim\sum (2l+1)/l^{3/2}e^{-\alpha_0 l}$ for $\alpha_0 l\gtrsim 1$.  Up to double-log corrections the first sum for $L<l\lesssim\alpha_0^{-1}$  can be estimated as $-2\log(\alpha_0 L)$ while the second one gives an order one constant $\gamma$. Eventually we find 
\bea
\label{FBS}
\sigma_{\rm total}\le  \sigma_0 Y^2(1-2\log(Y)+\gamma)\ ,\quad Y=\alpha_0{\mathcal A}(s)^{1/2}\ll 1\ .
\eea
The form of \eqref{FBS} is different from the canonical ``log-squared'' form \eqref{Froissart}. If ${\mathcal A}(s)$ grows polynomially, so is $Y$, while \eqref{FBS} is a combination of polynomial and logarithmic growth. For example, taking ${\mathcal A}(s)\sim s^2$ we get $Y\sim \sqrt{s}$ and \eqref{FBS} will be growing as $s \log(s)$. 
We have to  conclude there is no natural reason for \eqref{FBS} to grow as a very small power of $s$ reproducing \eqref{regge} or exhibiting a similar behavior.

Let us now discuss the area of validity of \eqref{fb} and \eqref{FBS}. The bound \eqref{FBS} is valid when $Y\ll 1$ and consequently when \eqref{FBS} is much smaller than $\sigma_0$. The bound \eqref{fb} is valid when $1/Y$ corrections are small compared with $(Y+1)^2$, i.e.~approximately starting from $Y\gtrsim 1$, and correspondingly when \eqref{fb} is larger than $\sim 2\sigma_0$. Obviously validity of \eqref{fb} improves with growth of $Y$.  When $Y\sim 1$ the bound on  the  total cross section is of order $\sigma_0$.

\subsection{Froissart Bound in Chiral Limit}
One line of thought suggests the Froissart-Martin bound \eqref{FM} is giving unrealistically high values for proton-proton total cross-section because pions, responsible for the lightest state appearing in $t$-channel, are pseudo-Goldstone bosons. Taking this logic to extreme it would be interesting to see what happens with the Froissart bound in chiral limit $m_\pi\rightarrow 0$. Naively in such a case $\sigma_0\sim 1/t_0$ becomes infinite and Froissart bound becomes obsolete.  Some authors suggested there is another improved bound  of the form \eqref{Froissart} with finite value of $\sigma_0$. Since large impact parameters correspond to a small invariant mass of the pion pair, and low energy Goldstone bosons decouple, it was argued in \cite{KS} that $\sigma_0$ would be determined by the strong interaction scale of $\sim$1 GeV. A similar conclusion was also reached in \cite{Giordano:2013iga,Giordano:2014cia,Greynat:2013cta,Greynat:2013zsa}.

Below we consider Froissart bound in QCD in chiral limit but reach different conclusions.  
Because of the Goldstone mechanism origin, at low energies  pions interact through derivative couplings. Hence for small physical $s$ and $t$ amplitude $A$ is small,  $A\rightarrow 0$ when $s,t\rightarrow 0$.  By extrapolating this property to unphysical region, we assume the jump on the branch-cut $\rho$ vanishes when $t\rightarrow 0$ and $s$ is large an physical.  In other words, we assume $|\rho(s,z)|\lesssim {\mathcal A}(s)(z-1)^a$ for some $a\ge 1$ at the vicinity of $z\rightarrow 1^+$. 

The following analysis is essentially the same for any $a$ and for convenience we fix $a=1$ unless noted otherwise. 
Using \eqref{z01} and \eqref{tail} we define
\bea\nonumber
\a_l&\equiv&{\mathcal A}(s)\int_{z_0}^\infty dz (z-1) Q_l(z)\lesssim {\mathcal A}(s)
\left\{
\begin{array}{l}
{2\over l^4}\qquad \ \, \quad   \qquad\qquad \, \ \ ,\quad  z_0-1\ll {25\over 8l^2} \\
{\sqrt{\pi} (1-e^{-\alpha_0})^{2+1/2}\over  2\sqrt{2}\, l^{3/2}}e^{-\alpha_0 l}\ ,\quad z_0-1\gtrsim \   {25\over 8l^2}
\end{array}\right. \\ & \label{eq:whatever}
\eea
 such that $|a_l(s)|< \a_l(s)$. Up to an $l$-independent pre-factor, for $l^2(z_0-1)\gg 1$ integral \eqref{eq:whatever}  is the same as \eqref{estimate} for any $a$. Hence, when ${\mathcal A}(s)$ is large enough and $Y\sim \log{\mathcal A}(s)\gtrsim 1$ the upper bound will be also given by \eqref{fb}. When $Y\ll 1$ the situation is different. In this case $L=2^{-1/4}{\mathcal A}^{1/4}(s)$ (more generally $L\sim{\mathcal A}^{1/(2a+2)}(s)$) and as before we split the sum \eqref{partialwaves} into two parts. First part, from $l=0$ to $L-1$ yields $L^2$, while the second part from $l=L$ to infinity is bound by $\sim \sum_{l\ge L} (2l+1)/l^4$ and converges to $\sim 1/L^2$. Hence 
\bea
\label{poly}
\sigma_{\rm total}\lesssim 2\sigma_0 Y^2 \ ,\quad Y= 2^{-1/4}\alpha_0{\mathcal A}(s)^{1/4}\ll 1\ .
\eea
For general $a$, $\sigma_{\rm total}\lesssim \sigma_0 Y^2(1+{1\over a})$ and $Y\sim \alpha_0{\mathcal A}(s)^{1/2(1+a)}$. 

Once again we notice that the asymptotic form \eqref{fb} corresponding to $Y\gtrsim 1$ is valid when $\sigma \gtrsim 2\sigma_0$ while the asymptotic form  corresponding to $Y\lesssim 1$ is valid when $\sigma \lesssim 2\sigma_0$. 

The bound \eqref{poly} remains finite in chiral limit $t_0\rightarrow 0$, 
\bea
\label{powerlaw}
\sigma_{\rm total} < c{{\mathcal A}(s)^{1/(1+a)} \over s}\ ,
\eea
where $c$ is an appropriate numerical coefficient. Assuming ${\mathcal A}(s)\sim s^N$, $N>(1+a)$, we arrive at one of 
the main conclusions of this paper: in chiral limit when pions are massless or when the observed $\sigma\lesssim \sigma_0$, Froissart bound may have a ``power law'' rather than ``log squared'' form. The ``power law'' asymptotic form of $\sigma$ as a function of $s$ is corroborated by a perturbative consideration of QCD evolution equation \cite{Kovner:2001bh}, although some authors believe arguments based on gluon saturation should lead to the conventional ``log squared'' asymptotic \cite{Ferreiro:2002kv}. Even though the possibility of the power law growth of the bound is remarkable, it  can hardly explain experimentally observed slow growth of \eqref{regge} through saturation. Indeed, small power of $s$ in \eqref{powerlaw} would require either fine-tuning of $N$ and $a$ to some fractional values, or unrealistically large $a$.

\section{Summary}
This paper is devoted to the question whether Froissart bound can explain experimentally observed slow growth of hadron total cross sections at high energies, in particular   total cross-section of proton-proton collisions. To this end we revisited derivation of Froissart bound, in particular considered what happens with the bound in chiral limit, when pions become massless. By making an appropriate assumption about the behavior of  function $\rho$ (jump of the scattering amplitude across the cut in $t$-plane) we derived new universal expression for the Froissart bound \eqref{powerlaw} which should be valid in theories with (pseudo)-Goldstone bosons while the cross-section $\sigma\ll \sigma_0=\pi/t_0$.

There are several findings which we believe have universal nature relevant for any QFT and do not depend on technical assumptions about analytic structure of scattering amplitude, e.g.~validity of dispersion relation.  Applying these results toward hadron scattering lead us to several interesting conclusions. Below we summarize our results. 
\begin{itemize}
\item In theories with mass gap there are several different regimes corresponding to different functional forms of Froissart bound. If $t_0$ is the lightest state appearing in $t$-channel, with an exception of bound states and stable particles, and $\sigma_0=\pi/t_0$, and the total cross-section $\sigma$ is significantly larger than 
$2\sigma_0$,  Froissart bound assumes its conventional ``log squared'' form (\ref{Froissart},\ref{fb}), provided  scattering amplitude grows with energy as a polynomial ${\mathcal A}(s)\sim s^N$. When $\sigma_F\lesssim \sigma_0$, functional form of Froissart bound may vary, but the common element is the power law growth of the bound with energy when $\sigma_F\ll \sigma_0$. 

In short, $\sigma\gtrsim 2\sigma_0$ is an approximate validity condition of the bound $\sigma<\sigma_0 \log^2(s/s_0)$.

\item In theories with massless Goldstone bosons, e.g.~QCD in chiral limit, Froissart bound will grow with energy polynomially \eqref{powerlaw}. This is a hypothesis based on the assumption of the particular  behavior of  the branch-cut jump function $\rho$ when Goldstones are the only massless states present in  the theory. 

\item In cases when Froissart bound grows with energy polynomially or approximately polynomially, e.g.~in theories with (pseudo)-Goldstones  when the total cross-section is much smaller than $\sigma_0$, 
we are unaware of a mechanism which would favor small power of polynomial growth. In the context of proton-proton total cross-section we conclude that the approximate empirical power-law growth with small scaling exponent in the energy range $10^2\sqrt{s}<10^5$ GeV, \eqref{regge}, can not be explained by saturation of the modification of the Froissart-Martin bound in the regime $\sigma_{pp}\lesssim \sigma^{FM}_0=63.32$ mb.

\item We equally conclude that the approximate empirical ``log squared'' growth of the proton-proton total cross-section \eqref{log2} in the range $10^2\sqrt{s}<10^5$ GeV can not be explained by saturation of the Froissart bound \eqref{Froissart} with $\sigma_0$ tuned to be of order of strong scale $\sigma_0\sim 1/M^2$, $M\sim 1$ GeV, to match the coefficient in front of $\log^2(s)$ from \eqref{log2}. Indeed,  empirical expression \eqref{log2} includes $s$-independent constant term $34.71$ mb which is either dominant or of the same order as $0.2647\log^2(s/16)$ for all energies within the range $10^2<\sqrt{s}<10^5$ GeV. As a result the proton-proton total cross-section is of order $40$ mb or larger, for $\sqrt{s}>10$ GeV, i.e.~several orders of magnitude larger than the proposed value of $\sigma_0\sim 0.2647$ mb. Consequently, for all $\sqrt{s}>10$ conventional Froissart bound \eqref{Froissart} with $\sigma_0\sim 0.2647$ mb must apply and correspondingly $\sqrt{s_0}$ must be of order $20$ MeV of smaller. This is three orders of magnitude smaller than the corresponding coefficient from \eqref{log2}. As a result \eqref{log2} will be significantly smaller than \eqref{Froissart} for all $10^2<\sqrt{s}<10^5$ GeV, thus defying the whole idea that \eqref{log2} saturates \eqref{Froissart} in this range.  

\item Finally, we have to conclude that  experimentally observed slow growth behavior of $\sigma_{pp}$ for $10^2<\sqrt{s}<10^5$ GeV can not be attributed to saturation of Froissar bound. It is an open question what is the true asymptotic of  $\sigma_{pp}$ when $s\rightarrow \infty$. It is likely that both ``power law'' \eqref{regge} and ``log squared'' behavior \eqref{log2} fail to capture true asymptotic of $\sigma_{pp}$. While the ``power law'' will ultimately violate Froissart-Martin bound, there is also some circumstantial  evidence that ``log squared'' behavior of $\sigma_{pp}$ does not continue at very high energies \cite{Azimov, Menon}. The Froissart-Martin bound should apply once $\sigma_{pp}$ becomes significantly larger than  $2\sigma_0^{FM}\simeq 130$ mb, i.e.~at energies above $10^5-10^6$ GeV.  We do not know if $\sigma_{pp}$ will eventually saturate the bound, but we expect $\sigma_{pp}$ to start exhibiting its true asymptotic behavior starting from this scale. Our expectations is a result of dimensional analysis. For $\sqrt{s}>10^6$ GeV there would be no other scale which could potentially interfere  with the asymptotic behavior. 

To conclude, we admit that currently available fits \eqref{log2} and \eqref{regge} may fail to predict $\sigma_{pp}$ at high energies above $\sqrt{s}>10^6$ GeV and expect that $\sigma_{pp}$ will start exhibiting its true asymptotic behavior at or below this scale.  

\end{itemize}

\section*{Acknowledgments}
I would like to thank Slava Rychkov for bringing this topic to my attention. I am thankful to Ya.~I.~Azimov, D.~Kharzeev, and M.~Stephanov for helpful correspondence and 
gratefully acknowledge support from the grant RFBR 15-01-04217.

\section*{A. Legendre Functions of the Second Kind}\label{sec:appendix}
Here we collect some useful relations related to the Legendre functions of the second kind $Q_l(z)$. We define $Q_l(z)$ through Legendre polynomial $P_l$ as follows
\bea
Q_l(z)\equiv{1\over 2}\int_{-1}^1 dx\, {P_l(x)\over z-x}\ ,\quad z>1\ .
\eea 
To derive the asymptotic form $Q_l(z)$ for large $l$ we use the following integral representation 
\bea
\label{ql}
Q_l(z)={1\over 2}\int_{-\infty}^{\infty} {dt\over (z+\sqrt{z^2-1}\cosh(t))^{l+1}}\ ,\quad z>1\ ,
\eea
and apply Laplace method to get
\bea
\label{Qapx}
\lim_{l\rightarrow \infty} Q_l(z)\rightarrow \sqrt{\pi\over 2(l+1)}{1\over (z^2-1)^{1/4}(z+\sqrt{z^2-1})^{l+1/2}}\ .
\eea
This is an accurate approximation of $Q_l(z)$ except of a small vicinity of $z \rightarrow 1^+$.  

Now we would like to estimate $\int_{z_0}^\infty dz\,  Q_l(z)$ for different values of $z_0$. Using \eqref{Qapx} and a change of variables $z=\cosh\alpha$ we immediately find
\bea
\label{int}
\int_{z_0}^\infty dz\,  Q_l(z)\simeq \sqrt{\pi\over 2(l+1)} \int_{\alpha_0}^\infty d\alpha\, e^{-(l+1/2)\alpha}\sinh^{1/2}\alpha\ , 
\eea
where  $\alpha_0=\log(z_0+\sqrt{z_0^2-1})$. There are two distinct regimes, $z_0<{(2l+1)\over 2\sqrt{l(l+1)}}\simeq 1+{1\over 8l^2}$ and $z_0> {(2l+1)\over 2\sqrt{l(l+1)}}\simeq 1+{1\over 8l^2}$. In the former case the peak of integrand is inside the interval of integration and one can once again apply Laplace method to find $z_0$-independent asymptotic
\bea
\int_{z_0}^\infty dz\,  Q_l(z)\simeq {\pi \over 2\sqrt{e} }{1\over l^{3/4}(l+1)^{1+1/4}}\ ,\quad z_0-1\lesssim {1\over 8 l^2}\ ,
\eea   
which correctly 
reproduces $1/l^2$ behavior of the exact  value for $z_0=1$
\bea
\label{Il}
I_l^0=\int_{1}^\infty dz\,  Q_l(z)={1\over l(l+1)}\ .
\eea
When $(z_0-1)\gtrsim {1\over 8l^2}$ the integral \eqref{int} can be estimated from above as 
\bea
\int_{z_0}^\infty dz\,  Q_l(z)<{\sqrt{\pi }\sqrt{1-e^{-2\alpha_0}}\over 2 l\sqrt{l+1}}e^{-\alpha_0 l}\ , \quad z_0-1\gtrsim {1\over 8 l^2}\ .
\eea

Similarly we can estimate $\int_{z_0}^\infty dz\,  (z-1)^a Q_l(z)$ for some positive $a$. Using the same change of variables as in \eqref{int} we find 
\bea
\label{inta}
\int_{z_0}^\infty dz\,(z-1)^a  Q_l(z)\simeq \sqrt{\pi\over 2(l+1)} \int_{\alpha_0}^\infty d\alpha\, e^{-(l+1/2)\alpha} (2\sinh^2(\alpha/2))^a \sinh^{1/2}\alpha\ .
\eea
Again, there are two regimes. When $z_0-1 \lesssim {(1+4a)^2\over 8l^2}$ the peak of integrand of \eqref{inta} is within the area of integration. Applying Laplace method to \eqref{inta} yields
\bea
\int_{z_0}^\infty dz\, (z-1)^a  Q_l(z)\simeq {\pi (1+4a)^{1+2a}\over 2^{1+3a}e^{1/2+2a}l^{2(a+1)}}\  ,\quad z_0-1\lesssim {(1+4a)^2\over 8 l^2}\ .
\eea  
This result correctly captures $1/l^{2(a+1)}$ behavior of $z_0 \rightarrow 1$ limit which can be deduced for positive integer $a$ from \eqref{Il}
and the recursion relation 
\bea
\label{z01}
I_l^a=\int_{1}^\infty dz\,  (z-1)^a Q_l(z)\ ,\quad
I_l^{a+1}={2(a+1)^2\over l(l+1)-(a+2)(a+1)} I_l^{a}\ .
\eea
When $z_0-1 \gtrsim {(1+4a)^2\over 8l^2}$ the peak of integrand of \eqref{inta} is outside of the area of integration, and the integral can be estimated from above as follows
\bea
\label{tail}
\int_{z_0}^\infty dz\, (z-1)^a\,  Q_l(z)<{\sqrt{\pi }\sqrt{1-e^{-2\alpha_0}}\over 2 (l-a)\sqrt{l+1}}{\left(1-e^{-\alpha_0}\right)^{2a}\over 2^a} e^{-\alpha_0 l}\ , \quad z_0-1\gtrsim {(1+4a)^2\over 8 l^2}\ .
\eea


\end{document}